\documentclass[aps, prd, twocolumn, superscriptaddress, longbibliography, nofootinbib]{revtex4-2}
\usepackage[utf8]{inputenc}
\usepackage[dvipsnames]{xcolor}
\usepackage{graphicx}
\usepackage{bm}
\usepackage{amsmath}
\usepackage{siunitx}
\usepackage{hhline}
\usepackage{multirow}
\usepackage{tabularx}
\usepackage{comment}
\usepackage{amssymb} 
\usepackage{enumerate}
\usepackage{tensor}
\usepackage{mathrsfs}
\usepackage{hyperref}
\usepackage[normalem]{ulem}
\usepackage{accents}
\usepackage{scalerel}
\usepackage{braket}
\usepackage{mathtools} % to use, e.g., \coloneqq ...

\makeatletter
\newcommand*{\leqdef}{\mathrel{\rlap{%
			\raisebox{0.25ex}{$\m@th\cdot$}}%
		\raisebox{-0.25ex}{$\m@th\cdot$}}%
	=}
\newcommand*{\reqdef}{=\mathrel{\rlap{%
			\raisebox{0.25ex}{$\m@th\cdot$}}%
		\raisebox{-0.25ex}{$\m@th\cdot$}}
}
\makeatother

\definecolor{DarkGreen}{rgb}{.1, .5, .1}

\begin{document}

\title{Gravitational wave stochastic background in reduced Horndeski theories}

\author{João C. Lobato}

\affiliation{Universidade Federal do Rio de Janeiro,
	Instituto de F\'\i sica\\
	CEP 21941-972 Rio de Janeiro, RJ, Brazil}

\author{Isabela S. Matos}

\affiliation{Universidade Federal do Rio de Janeiro,
	Instituto de F\'\i sica\\
	CEP 21941-972 Rio de Janeiro, RJ, Brazil}
	
\affiliation{Département de Physique Théorique and Center for Astroparticle Physics,
Université de Genève, Quai E. Ansermet 24, CH-1211 Genève 4, Switzerland}

\author{Maurício O. Calvão}

\author{Ioav Waga}

\affiliation{Universidade Federal do Rio de Janeiro,
	Instituto de F\'\i sica\\
	CEP 21941-972 Rio de Janeiro, RJ, Brazil}
	
\begin{abstract}
    We generalize to reduced Horndeski theories of gravity, where gravitational waves (GWs) travel at the speed of light,  the expression of a statistically homogeneous and unpolarized stochastic gravitational wave background (SGWB) signal measured as the correlation between the individual signals detected by two interferometers in arbitrary configurations. We also discuss some results found in the literature regarding cosmological distances in modified theories, namely, the simultaneous validity of a duality distance relation for GW signals and of the coincidence between the gravitational wave luminosity distance, based on the energy flux, and the distance inferred from the wave amplitude. This discussion allows us to conclude that the spectral energy density per unit solid angle of an astrophysical SGWB signal has the same functional dependency with the luminosity of each emitting source as in General Relativity (GR). Using the generalized expression of the GW energy-momentum tensor and the modified propagation law for the tensor modes, we conclude that the energy density of a SGWB maintains the same functional relation with the scale factor as in GR, provided that the modified theory coincides with GR in a given hypersurface of constant time. However, the relation between the detected signal and the spectral energy density is changed by the global factor $G_4(\varphi(t_0))$, thus potentially serving as a probe for modified gravity theories. 
\end{abstract}

\maketitle

\section{Introduction}

The first detection of gravitational waves (GWs) \cite{Abbott2016} opened a new window for astrophysical and cosmological inquiry. Almost a decade later from this astonishing experimental accomplishment, over 90 individual events were already observed \cite{Abbott2021}. 

But beneath these  well resolved detected signals, one expects to find a great number of faint unresolved gravitational waves originating from both astrophysical and cosmological sources. Although they cannot be observed separately, the effective signal resulting from their blend can be modelled, either because of a possible intrisically random process regarding their production or because of our lack of sufficient knowledge regarding their source distribution, as a stochastic signal, the stochastic gravitational wave background (SGWB) \cite{Romano2017, Caprini2018, Regimbau2011}, in some aspects similar to the electromagnetic (EM) and neutrino backgrounds. 

The Laser Interferometer Space Antenna (LISA) \cite{Whitepaper2017, Whitepaper2020}, a space-based GW detector expected to start operating in a decade from now, will provide the first millihertz survey of GWs, sensitive to signals with frequencies between $10^{-4}$ Hz and 1 Hz. LISA will be the first detector with a chance of getting a glance on the SGWB in general and, in particular, it is guaranteed to detect the background resulting from compact white dwarf binaries in our own galaxy \cite{Romano2017}.
This has  motivated the  study of anisotropies in the SGWB \cite{Cusin2020, Pitrou2020, Bartolo2020, Bartolo2022, Cusin2022} caused mainly by density galaxy contrasts, but also by Doppler/aberration effects due to peculiar velocities and other possible gravitational effects such as lensing \cite{Capurri2021}. 

The main objective of the present paper is to study possible generalizations of the SGWB signal in the context of the family of reduced Horndeski theories of gravity, that is, the Horndeski theories in which the GWs propagate at the speed of light. Lovelock has proven \cite{Lovelock1971, Lovelock1972} that the most general four dimensional theory with a Langrangian depending only on the metric whose field equations are second order differential ones is General Relativity (GR) with the presence of a cosmological constant. Maintaining the dimensionality and the order of the field equations, since introducing higher derivatives results in instabilities in the case of a nondegenerate Lagrangian \cite{Kobayashi2019}, the simplest way of generalizing Einstein's theory is to introduce a new scalar degree of freedom on the Lagrangian. The Horndeski theories \cite{Horndeski1974, Kobayashi2019}, derived in 1974, are the most general family of scalar-tensor theories in this context and gained more attention since 2012, when revisited by \cite{Charmousis2012}.

Possible  signatures of modified gravity in the SGWB signal in a reduced Horndeski theory could, in principle, originate from a modification of the GW energy-momentum tensor (EMT) when expressed as a function of the metric tensorial perturbation, the evolution of the GW amplitude over cosmological distances or even if the GW distance duality relation turned out not to be valid. This last aspect could affect SGWB signals coming from astrophysical sources since, when expressed in terms of the luminosity of the sources, their spectral energy density depends on the ratio between angular diameter and luminosity distances of the emitting galaxies. In view of these points, we shall analyze in the spirit of \cite{Isi2018}, if the signal is modified when compared with the one obtained in GR.

A discussion regarding the conceptually different cosmological distances will be made with the objective to attest the validity of the GW distance duality relation in the reduced Horndeski theories, even though it appears to be at odds with another result found in literature, \textit{i.e.} the coincidence between the gravitational wave luminosity distance and the distance inferred by the amplitude of GWs.

The paper is organized as follows. In Sec. \ref{sec: framework} we introduce the reduced Horndeski theories and alert for some simplifications assumed in our treatment. Additionally, we present the three relevant scales to our development and relate their order of magnitude. In Sec. \ref{sec: distances} we discuss the validity of the distance duality relation for GWs in the modified theory we are interested in and point out the conflict between it and another result found in literature. In Sec. \ref{sec:energy_density} the EMT and energy density of GWs for a general wave packet and for a statistically homogeneous and unpolarized SGWB are found as a function of the background scale factor in the context of a reduced Horndeski theory. In Sec. \ref{sec: signal} we derive the measured signal of the SGWB from the correlation between the individual perturbations in a pair of not coaligned and not coincident GW detectors in a Horndeski theory and find a non-trivial modification when compared with GR. Finally, we discuss how the validity of the GW distance duality relation on modified theories of gravity implies on no modification for the expression of the spectral energy density per unit solid angle as a function of the luminosities of galaxies hosting the sources of the astrophysical SGWB.

\section{Framework} \label{sec: framework}

The metric used in this paper consists of a cosmological background  perturbed by a GW,
\begin{align}
    g_{\mu\nu} = \bar{g}_{\mu \nu} + \epsilon h_{\mu \nu}, \label{metric}
\end{align}
where $\bar{g}_{\mu \nu}$ stands for the Friedmann-Lemaitre-Robertson-Walker (FLRW) metric
\begin{align}
    ds^2 = a^2(\eta) (-d\eta^2 + \delta_{ij}dx^idx^j),
\end{align}
with $\eta$ being the conformal time, $a(\eta)$ the scale factor and the real and dimensionless parameter $\epsilon$ being small, that is, $\epsilon\ll 1$. We will work in units in which $c=$ $\hbar=1$.

\subsection{Reduced Horndeski theories and simplifications} \label{subsec: Horndeski}

The modern way of expressing the Horndeski action is in its generalized Galileon form, which was proven in \cite{Kobayashi2011} to be equivalent to the original action derived for the theory. It is given by 
\begin{align}
	S_g \leqdef \frac{M_{Pl}^2}{2}\int\sqrt{-g} \sum_{i=2}^5 L_i d^4x. \label{complete_lagrangian}
\end{align}
where the Planck mass is just $M_{Pl}\coloneqq 1/\sqrt{8\pi G_{\textrm{N}}}$, with $G_{\textrm{N}}$ being Newton's gravitational constant, and
\begin{align}
	&L_2 \leqdef G_2(\varphi,X),\\
	&L_3 \leqdef G_3(\varphi,X) \Box \varphi, \\
	&L_4 \leqdef G_4(\varphi,X) R + G_{4,X}(\varphi,X) [(\Box \varphi)^2 - \varphi_{;\mu\nu}\varphi^{;\mu\nu}],\\
	&L_5 \leqdef G_5(\varphi,X)G^{\mu \nu} \varphi_{;\mu\nu} - \frac{1}{6}G_{5,X}(\varphi,X)[(\Box \varphi)^3 \nonumber \\ &\hspace{90pt}- 3 \varphi^{;\mu\nu}\varphi_{;\mu\nu}\Box\varphi+2\varphi_{;\mu}^{\;\;\;\nu}\varphi_{;\nu}^{\;\;\;\rho}\varphi_{;\rho}^{\;\;\;\mu}],
\end{align}
where $\varphi$ is the new scalar degree of freedom, ``," denotes simple partial derivative, ``;" indicates covariant ones and $\Box \varphi \leqdef g^{\mu\nu} \varphi_{;\mu \nu}$. Here, $X \leqdef - \varphi^{,\mu}\varphi_{,\mu}/2$ is the kinetic energy of the scalar field and $G^{\mu \nu}$ is the Einstein tensor. Note that, because a factor $M_{Pl}^2/2$ is already multiplying the $L_i$ quantities, GR is the special case in which $G_2 = -2\Lambda$, $G_3 = G_5 = 0$ and $G_4 = 1$. 

In 2017, the GW signal of a neutron star binary together with the detection of its electromagnetic (EM) counterpart \cite{Abbott2017, Goldstein2017} determined that GWs travel with the speed of light at low redshifts with an error of $10^{-15}$. With reasonable assumptions, this restricts Eq.~(\ref{complete_lagrangian}) to the reduced Horndeski theories, those in which \cite{Baker2017, Creminelli2017, Ezquiaga2017, Sakstein2017}
\begin{align}
	G_{4,X} = G_5 = 0 \label{speed_constraint}
\end{align} 
(see however \cite{Rham}). Then, the action becomes:
\begin{align}
	S_g = \frac{M_{Pl}^2}{2}\int\sqrt{-g} (G_2 + G_3 \Box \varphi + G_4 R) d^4x.
\end{align}

When studying GWs in Horndeski theories in the most general case, there are two elements that deserve special attention, as discussed in  \cite{Dalang2020,Dalang2021}. The first one is the assumption that not only the metric is perturbed, but that the scalar field is as well. In this case, one can show that such scalar perturbation behaves as a GW since it obeys a wave equation and introduces strains between freely-falling particles by means of a non-vanishing contribution to the Riemann tensor. Its evolution equation is coupled with the tensorial part $h_{\mu \nu}$. The second aspect regards the presence of a longitudinal mode in the tensorial perturbation, which, different from the GR case, cannot be eliminated globally. In other words, the transverse-traceless (TT) gauge can only be obtained locally. Although this additional mode is present, it does not contribute to the curvature of space-time and to the GW EMT, at least in the eikonal regime. To simplify our discussion, we will neglect both of these elements along all of our work. 

\subsection{Relevant Scales} \label{subsec: scales}

Here we briefly define three relevant scales following the formalism of \cite{Gravitation}. The first one is the typical variation of the background space-time: 
\begin{align}
    \mathcal{O}\left(\frac{1}{\mathcal{R}}\right) = \mathcal{O}\left(\frac{\bar{g}_{\alpha \beta, \mu}}{\bar{g}_{\gamma \delta}}\right) = \mathcal{O}(\mathcal{H})
\end{align}
where 
\begin{align}
    \mathcal{H} \leqdef \frac{a'}{a},
\end{align}
with $' \leqdef d/d\eta$. A second scale of variation is the one related  to the GWs: 
\begin{align}
    \mathcal{O}\left(\frac{1}{\Lambda}\right) = \mathcal{O}\left(\frac{h_{\alpha \beta, \mu}}{h_{\gamma \delta}}\right).
\end{align}

In order to distinguish the wave from the background, these two scales need to be of different order. Here we will choose to work in the sub-horizon regime, where 
\begin{align}
    \Lambda \ll \mathcal{R}. \label{subhorizon}
\end{align}
In this regime, several approximations can be made. As an example, since
\begin{align}
    \mathcal{O}\left(\bar{\Gamma}^{\alpha}_{\mu \nu}h_{\alpha \beta}\right) =  \mathcal{O}\left(\epsilon \mathcal{H}\right) = \mathcal{O}\left(\frac{\epsilon}{\mathcal{R}}\right),
\end{align}
where $\bar{\Gamma}^{\alpha}_{\mu \nu}$ is the background connection, one can conclude: 
\begin{align}
	h_{\mu \nu;\alpha} = h_{\mu \nu,\alpha} - \bar{\Gamma}^{\beta}_{\mu \alpha} h_{\beta \nu} - \bar{\Gamma}^{\beta}_{\alpha \nu} h_{\mu \beta} = h_{\mu \nu,\alpha}, \label{covariant_to_partial}
\end{align}
where from now on covariant derivatives and the raising and lowering of indexes will be understood as operations made with the background metric.

A final scale is related with the order of variation of the effective Planck mass $M_{Pl}^{eff}$. In the reduced Horndeski theories we are interested in, $M_{Pl}^{eff} = \sqrt{2 G_4}$ \cite{Bellini2014} and the related scale is:
\begin{align}
    	\mathcal{O}\bigg(\frac{1}{\mathcal{M}} \bigg) &= \mathcal{O}\bigg(\frac{(M_{Pl}^{eff})'}{M_{Pl}^{eff}}\bigg) = \mathcal{O}\bigg(\frac{G_4'}{G_4}\bigg). \label{eff_plank_mass_variation}
\end{align}
Since we are working with a background $\varphi$ and $G_{4,X} = 0$, the function $G_4$ is a background quantity and so we expect its variation to be no more than that of the background itself
\begin{align}
    \mathcal{O}\bigg(\frac{1}{\mathcal{M}} \bigg) \leq \mathcal{O}\bigg(\frac{1}{\mathcal{R}} \bigg). \label{M_greater_then_R}
\end{align}

\section{Distances with gravitational waves} \label{sec: distances}

Since the validity of the duality relation between GW cosmological distances  will be important to the following discussion of the SGWB signal, we take a moment to dwell on the conceptually distinct distances in Horndeski theories. 

There are at least 5 different concepts of distance one can adopt in the present context. The first two are related with angular measurements and the size of the source emitting GWs or electromagnetic waves (EMWs). They are called GW/EM angular size (or diameter) distances, defined as
\begin{align}
    D_A^{(EM,GW)} \leqdef \sqrt{\frac{\delta A_s^{(EM,GW)}}{\delta \Omega_o^{(EM,GW)}}}, \label{diameter_distance}
\end{align}
where $\delta A_s^{(EM,GW)}$ is the (infinitesimal) area of the source as measured  by an instantaneous observer at it and $\delta \Omega_o^{(EM,GW)}$ is the (infinitesimal) solid angle subtended by the source as measured by an instantaneous observer at a detector. We stress that, even if the EMWs and GWs from the source arise from distinct areas, insofar as the limit implicit in Eq. (\ref{diameter_distance}) is well defined, both angular distances will coincide: because both photons and gravitons travel along null geodesics, one can use a purely geometric argument \cite{Ellis2012}, for any field equation governing gravity that has the FLRW metric as a solution, to conclude
\begin{align}
    D_A^{(EM,GW)} = a(t_e)\int^{t_o}_{t=t_e} \frac{dt}{a(t)}.
\end{align}
where $t_e$ and $t_o$ are the cosmic times of emission and observation, respectively. Thus, if GWs and EMWs are being emitted radially and at the same event by the source, they will arrive here together as well, because of the uniqueness of a geodesic curve  given a initial condition,  and we have no reason to believe these two distances differ:
\begin{align}
    D_A^{(EM)} = D_A^{(GW)} = D_A. \label{angular_distances}
\end{align}

Another pair of distances, the luminosity distances, are related with the energy flux. Their definition is:
\begin{align}
    D^{(EM,GW)}_L \leqdef \sqrt{\frac{L^{(EM,GW)}}{4 \pi \phi^{(EM,GW)}}} 
\end{align}
where $L^{(EM,GW)}$ is the power emitted by the source near it (also called the luminosity of the source) and $\phi^{(EM,GW)}$ is the energy flux measured here. In this case, it is not as obvious that the two distances coincide, since the EM and the GW EMTs are different from one another.

The last distance, which we shall call the GW amplitude distance (also called the gravitational distance in \cite{Dalang2020}), can be expressed as:
\begin{align}
    D_h^{(GW)} = \sqrt{\frac{G_4(\varphi_o)}{G_4(\varphi_e)}} D_L^{(EM)}. \label{D_amp}
\end{align}
where $\varphi_e$ and $\varphi_o$ are the values of $\varphi$ in emission and observation events. Its motivation comes from the fact that the amplitude of a GW emitted by a binary system falls with $D_L^{(EM)}$in GR, while in a reduced Horndeski theory it falls with $D_h^{(GW)}$. This change in the GW propagation through cosmological distances can be used to probe modified gravity \cite{Amendola2018,Belgacem2019,Lagos2019,Matos2021}. It is common to find works \cite{Belgacem2019, Tasinato2021} referring to $D_h^{(GW)}$ as the GW luminosity distance, but here we distinguish it from $D^{(GW)}_L$, which is the distance directly related with the luminosity of the source, since in general  $D_h^{(GW)} \neq D^{(GW)}_L$, as is shown further below. 

We now point out three results found in literature that, in light of what was previously presented, lead to a contradiction. This discussion is intimately related with the topic of this paper since one of the three results will be relevant for the SGWB calculations.

The first result is the standard distance duality relation for EM signals, valid in any pseudo-Riemannian theory of gravity in which photons travel along null geodesics and have their number conserved \cite{Etherington1933, Ellis2007, Santana2017}:
\begin{align}
    D_L^{(EM)} = (1+z)^2 D_A^{(EM)}. \label{EM_duality}
\end{align}
The second result is an analogous relation for the corresponding gravitational distances, that is 
\begin{align}
    D_L^{(GW)} = (1+z)^2 D_A^{(GW)}. \label{GW_duality}
\end{align}
This relation was proven to be true in \cite{Tasinato2021} for a general scalar-tensor theory and a GW signal of the eikonal form
\begin{align}
h_{\alpha \beta} = Re\{H_{\alpha \beta}(x^{\mu}) e^{iw(x^{\mu})} \} \label{eikonal}
\end{align}
where $w$ is a real function, $H_{\alpha \beta}$ is a complex one and $Re$ denotes the real part of a function. The third and last result is found on Appendix A of \cite{Belgacem2019}, where the authors claim to prove (compare Eqs. (2.16) and (A.11) of their work) that $D_h^{(GW)} = D_L^{(GW)}$. 

The contradiction is that, once Eq.~(\ref{angular_distances}) is used on Eqs.~(\ref{EM_duality}) and (\ref{GW_duality}), one concludes that $D^{(EM)}_L = D^{(GW)}_L$, which, together with $D_h^{(GW)} = D_L^{(GW)}$ implies $D_h^{(GW)} = D^{(EM)}_L$, in disagreement with Eq.~(\ref{D_amp}). The way \cite{Tasinato2021} finds to escape this contradiction is to conclude that $D_A^{(GW)} \neq D_A^{(EM)}$, but, as was argued before, Eq.~({\ref{angular_distances}}) is a purely geometrical result, valid for all radial null geodesics, independent of the nature of the emitted particle (photon or graviton). The only two possible solutions to this problem seem to be the invalidation of Eq.~(\ref{GW_duality}) or the conclusion that $D_h^{(GW)}\neq D_L^{(GW)}$.

We argue now why Eq.~(\ref{GW_duality}) must be right. The GW EMT of the signal in Eq.~(\ref{eikonal}) in a reduced Horndeski theory is \cite{Dalang2020}:
\begin{align}
    T^{(GW)}_{\mu \nu} = \frac{1}{8} M_{Pl}^2 G_4 |H_{\perp}|^2 q_{\mu} q_{\nu}, \label{EMT_eikonal}
\end{align}
where $q_{\mu} \leqdef w_{,\mu}$ and $|H_{\perp}|^2\leqdef |H_{\circlearrowleft}|^2 + |H_{\circlearrowright}|^2$, with $H_{\circlearrowleft}$ and $H_{\circlearrowright}$ being the two circular polarizations of the GW. Defining the graviton flux density as
\begin{align}
    J_{\nu}^{(GW)} \leqdef -\frac{u^{\mu}T^{(GW)}_{\mu \nu}}{\omega_{GW}} = \frac{\rho_{GW}}{\omega_{GW}^2} q_{\nu}, \label{current}
\end{align}
where $\omega_{GW}\leqdef -u^{\alpha}q_{\alpha}$, $\rho_{GW} \leqdef u^{\mu} u^{\nu} T^{(GW)}_{\mu \nu}$ and $u^{\alpha}$ is the instantaneous observer measuring these quantities, it is possible to use the  field equations of the modified theory of gravity to prove that the EMT of Eq.~(\ref{EMT_eikonal}) obeys the continuity law \cite{Dalang2020, Tasinato2021}
\begin{align}
    J_{\nu}^{(GW);\nu} = 0, \label{divergenceless}
\end{align}
which is equivalent to $T_{\mu \nu}^{(GW);\nu} = 0$. Replacing Eq.~(\ref{current}) in Eq.~(\ref{divergenceless}) one finds
\begin{align}
    \frac{d}{d\vartheta}\left(\frac{\rho_{GW}}{\omega_{GW}^2}\right) + \hat{\Theta}\frac{\rho_{GW}}{\omega_{GW}^2} = 0. \label{energy_evol}
\end{align}
where $\hat{\Theta} \leqdef q_{;\mu}^{\mu}$ and $d/d\vartheta \leqdef q^{\mu} \nabla_{\mu}$. We remember that $\hat{\Theta}$ measures the expansion/contraction of the cross-sectional area of the graviton beam and can be expressed as \cite{Ellis2012} (notice the typo in Eq.(2.75) of the reference given):
\begin{align}
    \hat{\Theta} = \frac{1}{\delta S}\frac{d \delta S}{d\vartheta}, \label{cross_section_expansion}
\end{align}
where $\delta S$ is the cross-sectional area of the beam. With Eq.~(\ref{cross_section_expansion}), one can integrate Eq.~(\ref{energy_evol}) to conclude
\begin{align}
    \frac{\rho_{GW}}{\omega_{GW}^2} \delta S = constant. \label{number_conservation}
\end{align}
Since $2\pi\rho_{GW}/\omega_{GW}$ is the number density of gravitons, this result can be interpreted as stating the conservation of the number of gravitons within each volume $\lambda_{GW} \delta S$ along the null geodesics. 

Starting from Eq.~(\ref{number_conservation}), it is possible to conclude Eq.~(\ref{GW_duality}) in a way analogous to the EM case. For the sake of completeness, we present how the argument to that end develops. Assume the infinitesimal bundle of gravitons are emitted at event E and observed at event O,  Eq.~(\ref{number_conservation}) implies that
\begin{align}
    \phi^{(GW)}|_E \delta S|_E &=  \left(\frac{\lambda_{(GW)}|_O}{\lambda_{(GW)}|_E}\right)^2 \phi^{(GW)}|_O \delta S|_O \nonumber \\ &= (1+z)^2   \phi^{(GW)}|_O \delta S|_O \label{emitted_and_observed_fluxes}
\end{align}
where we used that the GW flux coincides with the energy density, since we are assuming $c=1$, and, in the last step the definition of redshift was used. Now, the flux at emission multiplied by the area $\delta S_E $ and a time interval $\delta t_e$, gives the energy emitted during that period and in that area. We rewrite this energy in terms of the luminosity of the source per solid angle ($L^{(GW)}_{\Omega}$):
\begin{align}
    \phi^{(GW)}|_E \delta S|_E  \delta t|_E  \leqdef \delta E|_E \reqdef L^{(GW)}_{\Omega}|_E \delta \Omega|_E \delta t|_E.
\end{align}
This allows us to express Eq.~(\ref{emitted_and_observed_fluxes}) as
\begin{align}
    \frac{L^{(GW)}_{\Omega}|_E}{\phi^{(GW)}|_O}  = (1+z)^2 \frac{\delta S|_O}{\delta \Omega|_E} \label{almost_duality}
\end{align}
Assuming a source emitting GWs isotropically, the total luminosity will be just $L^{(GW)} = 4 \pi L^{(GW)}_{\Omega}$ and the left-hand-side of the above equation is, then, the square of the GW luminosity distance. As for the ratio between the infinitesimal quantities in the right-hand-side, it defines a similar concept of distance as the diameter distance of Eq.~(\ref{diameter_distance}). The difference is that $D_A$ is constructed imagining a bundle diverging from O with solid angle $\delta \Omega|_O$ and reaching E with a certain cross-sectional area $\delta S|_E$, while in the present case the opposite occurs, the bundle emerges from E with solid angle $\delta \Omega|_E$ and reaches O with area $\delta S|_O$. In this sense, we can define a reciprocal angular distance by
\begin{align}
    D_{rec}^{(GW)} \leqdef \sqrt{\frac{\delta S|_O}{\delta \Omega|_E}}. \label{reciprocal_distance}
\end{align}
A famous and important result is the relation between Eq.~(\ref{diameter_distance}) and Eq.~(\ref{reciprocal_distance}). It is called the reciprocity relation, valid for any bundle of null geodesics. It reads \cite{Ellis2007, Etherington1933, Santana2017}:
\begin{align}
    D_{rec}^{(GW)} = (1+z) D_{A}^{(GW)}. \label{reciprocity}
\end{align}
Inserting Eq.~(\ref{reciprocity}) in Eq.~(\ref{almost_duality}), we conclude the duality relation, Eq.~(\ref{GW_duality}).

We conclude, then, that it is safe to use Eq.~(\ref{GW_duality}) and that, consequently, $D_h^{(GW)} \neq D^{(GW)}_L$. 

\section{Energy density of a SGWB} \label{sec:energy_density}

We now want to investigate if there are possible functional changes in the expression giving the energy density of a SGWB in terms of the scale factor. This investigation is motivated by the fact that, in the reduced Horndeski theories, the functional expression of the GW EMT in terms of the $h_{ij}$ is modified by a factor $G_4$   \cite{Dalang2020}:
\begin{align}
    T^{(GW)}_{\mu \nu} = G_4 \frac{M_{Pl}^2}{4} \langle h_{i j;\mu} h^{i j}_{;\nu}  \rangle = G_4 \frac{M_{Pl}^2}{4} \langle h_{i j,\mu} h^{i j}_{,\nu}  \rangle \label{EMT_general}
\end{align}
where Eq.~(\ref{covariant_to_partial}) was used in the last step and $\langle ... \rangle$ stands for a space-time average under an intermediate scale, much greater than the GW variation scale $\Lambda$ but much smaller than the background cosmological one $\mathcal{R}$ \cite{Isaacson1968II, Burnett1989, Preston2016}. The correspondent GW energy density as measured by observers in the Hubble flow is
\begin{align}
    \rho_{GW} = G_4 \frac{M_{Pl}^2}{4} \Big\langle \dot{h}_{i j} \dot{h}^{i j}  \Big\rangle, \label{energy_density}
\end{align}
where $\dot{} \leqdef d/dt$. Although the $G_4$ factor alters the GW EMT functional dependence with the GW amplitude when compared with GR, one needs to notice that the $h_{ij}$ perturbations have their propagation modified as well in these more general theories. It is, thus, important to verify how these two effects combine, that is, to verify if the GW EMT has  a modified functional dependence when expressed in terms of the scale factor. 

In a reduced Horndeski theory, the propagation equation for the GW amplitude is \cite{Saltas2014}
\begin{equation}
\tilde{h}''_{ij} + 2\left(\mathcal{H}+\frac{G_4'}{2G_4}\right)\tilde{h}'_{ij} + k^2\tilde{h}_{ij} = 0.\label{gw_propagation}
\end{equation}
where 
\begin{align}
	\tilde{h}_{ij} (\eta,k_i) \leqdef \int h_{ij} (\eta, x^i) e^{-ik_i x^i} d^3x,
\end{align}
is the spatial Fourier transform of the perturbation, $k_i$ are real and event independent quantities and $k^2 \leqdef \delta^{i j} k_{i} k_{j}$. Implicitly defining the quantity $a_{eff}$ as
\begin{align}
    \frac{a_{eff}'}{a_{eff}} \leqdef \left(\mathcal{H} + \frac{G_4'}{2G_4}\right) \label{eff_expansion}
\end{align}
it is possible to arrive, under the subhorizon regime of Eq.~(\ref{subhorizon}), at a plane wave solution for $\tilde{h}_{ij}$ of the form \cite{Amendola2018, Belgacem2019}
\begin{align}
    \tilde{h}_{ij} = \frac{1}{a_{eff}} A_{ij}(k_i) e^{2\pi i f \eta},
\end{align}
provided the dispersion relation
\begin{align}
    k^2 = (2 \pi f)^2 \label{dispersion}
\end{align}
is satisfied. This implies the following  wave-packet signal
\begin{align}
  	h_{ij} = \frac{1}{a_{eff}(\eta)} \int  A_{ij}(k_i) e^{2\pi i f (\eta+\hat{\Omega}_i x^i)} d^3k, \label{wave-packet}
\end{align}
where $\hat{\Omega}_i \leqdef k_i/k$ and Eq.~(\ref{dispersion}) was used in the exponential argument. Since the SGWB is a combination of different signals coming from different sources, it is described by a wave-packet of this form, where the functions $A_{ij}$ are understood to be random variables.

In GR, Eq.~(\ref{gw_propagation}) becomes
\begin{align}
	(\tilde{h}_{ij}^{(GR)})'' + 2\mathcal{H}^{(GR)}(\tilde{h}_{ij}^{(GR)})' + k^2\tilde{h}_{ij}^{(GR)} = 0,
\end{align}
with the analogous solution being
\begin{align}
  	h^{(GR)}_{ij} = \frac{1}{a^{(GR)}(\eta)} \int  A_{ij}(k_i) e^{2\pi i f (\eta+\hat{\Omega}_i x^i)} d^3k, \label{wave_packet_GR}
\end{align}
where the scale factor in this case would obey a different field equation than the one in the modified theory, but we assume the amplitudes $A_{ij}$ to be the same as in the Horndeski theories. 

Differentiating Eq.~(\ref{wave-packet}) we arrive at
\begin{align}
    h_{ij,\nu} +  \delta^0_{\nu} \frac{a'_{eff}}{a_{eff}} h_{ij} = \frac{i}{a_{eff}} \int A_{ij} k_{\nu} e^{ 2\pi i f  (\eta+\hat{\Omega}_i x^i)} d^3 k, \label{derivative_of_h}
\end{align}
where $k_{\alpha} \leqdef 2 \pi f(1, \hat{\Omega}_i)$. Since $\mathcal{O}(h_{ij,\nu}) = \mathcal{O}(\epsilon/\Lambda)$ and, by Eqs.~(\ref{eff_plank_mass_variation}), (\ref{M_greater_then_R}) and (\ref{eff_expansion}), $\mathcal{O}(h_{ij}a_{eff}'/a_{eff}) = \mathcal{O}(\epsilon/\mathcal{R})$, we can safely neglect the second term of the left-hand side of Eq.~(\ref{derivative_of_h}) to conclude
\begin{align}
     h_{ij,\nu} = \frac{i}{a_{eff}} \int  A_{ij} k_{\nu} e^{ 2\pi i f  (\eta+\hat{\Omega}_i x^i)} d^3 k. \label{approx_derivative_of_h}
\end{align}

Replacing Eq.~(\ref{approx_derivative_of_h}) in Eq.~(\ref{EMT_general}), using that $h_{ij}$ is a real function and that $a_{eff}$ varies like the background so that it can leave the space-time average
\begin{align}
    T^{(GW)}_{\mu \nu} &= G_4 \frac{M_{Pl}^2}{4} \langle h_{i j,\mu} (h^{i j}_{,\nu})^*  \rangle \nonumber \\ &= G_4 \frac{M_{Pl}^2}{4 a^2_{eff}}  \Big\langle \int \int  A_{ij} (k_i) (A^{ij})^*(\tilde{k}_i)  \times \nonumber \\ &\hspace{20pt} \times k_{\mu} \tilde{k}_{\nu} e^{2\pi i[ (f-\tilde{f})\eta-(\tilde{f}\tilde{\hat{\Omega}}_i - f\hat{\Omega}_i) x^i]} d^3 k d^3\tilde{k}\Big\rangle, \label{TEM_simplified}
\end{align}
where $*$ denotes the complex conjugate of a function.  Integrating Eq.~(\ref{eff_expansion}), we find
\begin{align}
   a_{eff} = B a \sqrt{G_4}, \label{general_a_eff}
\end{align}
where $B$ is a constant. So we conclude that there is an additional new contribution to the EMT coming from the GW propagation law, namely, the factor $a_{eff}^{-2}$ that also depends on the function $G_4$, whereas in GR would only depend on $a$. We set a general initial condition for $a_{eff}$ so that $B$ can be determined. We assume that at a given time $t_i$, $a_{eff}$ assumes a value $a_{eff}(t_i)$. We then have
\begin{align}
a_{eff} (t) = a_{eff}(t_i) \frac{a(t)}{a(t_i)} \sqrt{\frac{G_4(t)}{G_4(t_i)}} \label{part_a_eff},
\end{align}
which implies in the following GW EMT:
\begin{align}
	 T^{(GW)}_{\mu \nu} &= Q(t_i)\frac{M_{Pl}^2}{4 a^2}  \Big\langle \int \int  A_{ij} (k_i) [A^{ij}(\tilde{k}_i)]^* \times \nonumber \\ &\hspace{15pt} \times k_{\mu} \tilde{k}_{\nu} e^{2\pi i[ (f-\tilde{f})\eta-(\tilde{f}\tilde{\hat{\Omega}}_i - f\hat{\Omega}_i) x^i]} d^3 k d^3\tilde{k}\Big\rangle. \label{EMT_wave_packet}
\end{align}
where
\begin{align}
    Q(t_i) &\leqdef G_4(t_i)  \left[\frac{a(t_i)}{a_{eff}(t_i)}\right]^2 \nonumber \\ &= G_4(t_i) \left[\frac{h_{ij}(t_i)}{h_{ij}^{(GR)}(t_i)}\right]^2 \left[\frac{a(t_i)}{a^{(GR)}(t_i)}\right]^2 \label{init_condit_param}
\end{align}
where we used Eqs.~(\ref{wave-packet}) and (\ref{wave_packet_GR}) in the last step. The function $Q(t_i)$ reunite all three initial conditions of the fields of the model considered: the background metric, the tensorial perturbation and the scalar field (present implicitly in the argument of $G_4$).
If we choose coincident initial conditions for $a$ and $h_{ij}$ in both theories, then $Q(t_i) = G_4(t_i)$. The possibility of setting $G_4(\varphi(t_i)) = 1$ will depend on the theory, since this can only be done if there is a value for $\varphi$ in which $G_4$ becomes 1.

The energy density is obtained by a similar procedure, but starting with Eq.~(\ref{energy_density}) and reads
\begin{align}
    	\rho_{GW} &= G_4 \frac{M_{Pl}^2}{4} \langle \dot{h}_{i j} (\dot{h}^{i j})^*  \rangle \nonumber \\ &= Q(t_i) \frac{M_{Pl}^2 \pi^2}{a^4}  \Big\langle \int \int  A_{ij} (k_i) (A^{ij})^*(\tilde{k}_i) \times \nonumber \\ &\hspace{20pt} \times f \tilde{f} e^{2\pi i[ (f-\tilde{f})\eta-(\tilde{f}\tilde{\hat{\Omega}}_i - f\hat{\Omega}_i) x^i]} d^3 k d^3\tilde{k}\Big\rangle. \label{energy_density_wave_packet}
\end{align}

Eqs.~({\ref{EMT_wave_packet}}) and (\ref{energy_density_wave_packet}) are valid for any wave-packet. We now specialize for the case of SGWB, where the $A_{ij}$ are treated as random variables. More precisely, expanding the $A_{ij}$ into the polarization basis\footnote{For a precise definition of the basis, see, for example, \cite{Allen99}.}:
\begin{align}
    A_{ij}(k_i) = A_P(k_i) e^P_{ij}(\hat{\Omega}), \label{polarizations}
\end{align}
where $P=+,\times$ indicates the GW polarizations, the SGWB is characterized by the random nature of the $A_P$. In this case, assuming ergodicity, one may exchange the average in space-time coordinates to an average in the random amplitude functions \cite{MaggioreVol1}.  The energy density becomes:
\begin{align}
    \rho_{GW} &=  Q(t_i) \frac{ M_{Pl}^2 \pi^2}{a^4}   \int \int  \langle A_{P} (k_i) (A_{\tilde{P}})^*(\tilde{k}_i) \rangle e^{P}_{ij}(\hat{\Omega}) \times \nonumber \\ &\hspace{10pt} \times e^{\tilde{P}ij}(\tilde{\hat{\Omega}})  f \tilde{f}  e^{2\pi i[ (f-\tilde{f})\eta-(\tilde{f}\tilde{\hat{\Omega}}_i - f\hat{\Omega}_i) x^i]} d^3 k d^3\tilde{k}. \label{energy_density_SGWB}
\end{align}

We assume the SGWB to be statistically homogeneous, that is, the mean value and the variance of the tensorial perturbation are independent of $x^i$. This implies:
\begin{align}
    \frac{\partial}{\partial x^m}\langle h_{ij} (x^{\alpha}) [h_{lk}(x^{\alpha})]^* \rangle = 0. 
\end{align}
Inserting the GW wave-packet of Eq.~(\ref{wave-packet}) in the above restriction, together with Eq.~(\ref{polarizations}) one concludes that 
\begin{align}
    \langle A_P(k_i) [A_{\tilde{P}}(\tilde{k}_i)]^* \rangle &= N_{P \tilde{P}}(k_i) \delta^{(3)}(k_i - \tilde{k}_i) \nonumber \\ &= N(f,\hat{\Omega}) \delta^{(3)}(k_i - \tilde{k}_i) \delta_{P \tilde{P}}, \label{hom_unp_SGWB}
\end{align}
where, in the last step, we also assumed unpolarized SGWB and used Eq.~(\ref{dispersion}) to write $N(k_i) = N(f, \hat{\Omega}_i)$. Notice that under these assumptions, the signal can still be anisotropic, since $N$ depends on $\hat{\Omega}$. We mantain this level of generality, since one expects anisotropies on the SGWB signal arising from boost effects between the solar system and the SGWB frames \cite{Cusin2022}. Because the signal of Eq.~(\ref{wave-packet}) must be real, it obeys $A_{ij}^*(f, \hat{\Omega}) = A_{ij}(-f, \hat{\Omega})$.
Making the tranformation $f \rightarrow -f$ and $\tilde{f} \rightarrow - \tilde{f}$ we get
\begin{align}
    \langle A_P(-f, \hat{\Omega}) A_{\tilde{P}}(\tilde{f}, \tilde{\hat{\Omega}}) \rangle &= N (-f, \hat{\Omega}) \delta^{(3)}(k_i - \tilde{k}_i) \delta_{P \tilde{P}}.
\end{align}
Then, exchanging variables by the rules $f \leftrightarrow \tilde{f}$, $\hat{\Omega} \leftrightarrow \tilde{\hat{\Omega}}$ and $P \leftrightarrow \tilde{P}$, we find
\begin{align}
    \langle A_{\tilde{P}}(-\tilde{f}, \tilde{\hat{\Omega}}) A_{P}(f, \hat{\Omega}) \rangle &= N (-\tilde{f}, \tilde{\hat{\Omega}}) \delta^{(3)}(k_i - \tilde{k}_i) \delta_{P \tilde{P}}\nonumber \\ &= N (-f, \hat{\Omega}) \delta^{(3)}(k_i - \tilde{k}_i) \delta_{P \tilde{P}}. \label{minus_hom_unp_SGWB}
\end{align}
where the last equality holds because of the delta factors. Comparing Eqs.~(\ref{hom_unp_SGWB}) and (\ref{minus_hom_unp_SGWB}), we conclude that
\begin{align}
    N(-f, \hat{\Omega}) = N(f, \hat{\Omega}),
\end{align}

Replacing Eq.~(\ref{hom_unp_SGWB}) in Eq.~(\ref{energy_density_SGWB}) and noting that \cite{MaggioreVol1}
\begin{align}
   \delta_{P \tilde{P}} e^P_{ij} e^{\tilde{P}ij} = 4,
\end{align}
the energy density becomes
\begin{align}
      \rho_{GW} &= Q(t_i) \frac{4 M_{Pl}^2 \pi^2}{a^4} \int   N(k_i)  f^2  d^3k. 
\end{align}
Finally, we use Eq.~(\ref{dispersion}) to write $d^3k = f^2 df d^2\Omega$ and the parity in the first argument of the function $N$ to obtain
\begin{align}
      \rho_{GW} &= Q(t_i) \frac{8 M_{Pl}^2 \pi^2}{a^4} \int_{S^2} \int_{0}^{\infty}   N(f, \hat{\Omega})  f^4  df d^2 \Omega, \label{final_energy_density_SGWB}
\end{align}
where $S^2$ is the two sphere. We notice that the usual $a^{-4}$ behavior of a radiation fluid is recovered for the energy density of a SGWB. The function $Q(t_i)$ is simply unity in GR, since, in this case, $G_4(\varphi(t_i)) = 1$. We can recover this value for $Q(t_i)$ even in Horndeski theories if we assume that there is a hypersurface of constant $t$, at early times for instance, so that the modified theory coincides with GR.   
In the next section, we relate the energy density with the SGWB measured signal.

\section{Observable signal} \label{sec: signal}

\subsection{The signal}

We here will adapt the rationale in \cite{Allen99} to obtain the expression for the observable signal of SGWB, but now in the Horndeski theory. Assume two GW detectors in different (spatial) positions and with aribtrarily oriented arms. Their geometrical configuration is characterized by the tensor
\begin{align}
    D^{ij}_l \leqdef \frac{\hat{X}^i_l\hat{X}^j_l - \hat{Y}^i_l\hat{Y}^j_l}{2},
\end{align}
where $l=1,2$ indexes the two possible detectors and $\hat{X}^i_l$ and $\hat{Y}^i_l$ give the directions of the two arms of the $l$-th detector.
The measured signal in each detector will be:
\begin{align}
    s_l \leqdef h_l + n_l,
\end{align}
where
\begin{align}
    h_l \leqdef h_{ij}D^{ij}_l
\end{align}
and $n_l$ denotes the noise. The observable signal is\footnote{Here, for simplifying purposes, we choose the filtering function to be $\delta(\eta_1 - \eta_2)$, since we do not believe that choosing the optimal filtering will result in additional corrections from the GR case.} 
\begin{align}
    \langle S \rangle &\leqdef \int^{t_0 + T/2}_{t_0 -T/2} \langle s_1(\eta(t), x^i_1) s_2(\eta(t), x^i_2) \rangle dt \nonumber \\ &= \int^{t_0 +T/2}_{t_0 -T/2} \langle h_1(\eta(t), x^i_1) h_2^*(\eta(t), x^i_2) \rangle dt,
\end{align}
where $x^i_l$ is the position of the $l$-th detector, $T\approx 10$ years is the typical period of observation and in the last step we used that $h_2 = h_2^*$ and assumed the noises to be uncorrelated:
\begin{align}
    \langle n_1 n_2 \rangle = \langle n_i h_j \rangle = 0
\end{align}
which is a valid assumption if the detectors are sufficiently far apart.

Substituting now Eq.~(\ref{wave-packet}) and evaluating the mean with Eq.~(\ref{hom_unp_SGWB}), the signal becomes
\begin{align}
    \langle S \rangle = \int^{t_0+T/2}_{t_0-T/2} \frac{1}{a_{eff}^2} \int_{S^2} \int^{\infty}_{-\infty} N(f, \hat{\Omega}) F^P_1(\hat{\Omega}) F_{P2}(\hat{\Omega}) \times \nonumber \\ &\hspace{-140pt} \times  e^{2\pi i f \hat{\Omega}_i (x^i_1 - x^i_2)} f^2 df d^2 \Omega dt, \label{product_signal}
\end{align}
where
\begin{align}
    F_l^P \leqdef D^{ij}_l e^P_{ij}
\end{align}
and
\begin{align}
    F_1^P F_{P2} \leqdef \delta_{P \tilde{P}} F_1^P F_{2}^{\tilde{P}}
\end{align}

To relate the signal with the energy density, we define the latter in terms of the spectral energy density per unit solid angle as 
\begin{align}
    \rho_{GW} \leqdef \int_{S^2} \int_0^{\infty} \frac{d^3 \rho_{GW}}{df d^2 \Omega} df d^2\Omega
\end{align}
this equation must be valid together with Eq.~(\ref{final_energy_density_SGWB}) for any $N(f,\hat{\Omega})$ since the amplitudes $A_{ij}(k_i)$ in Eq.(\ref{hom_unp_SGWB}) are arbitrary. Then, one must have
\begin{align}
    N(f, \hat{\Omega}) = \frac{1}{Q(t_i)}\frac{a^4 }{8M_{Pl}^2\pi^2f^4} \frac{d^3 \rho_{GW}}{df d^2 \Omega}. \label{function_N}
\end{align}
Inserting this function on Eq.~(\ref{product_signal}) together with Eq.~(\ref{part_a_eff}), the signal reads
\begin{align}
      \langle S \rangle = \frac{G_{N}}{\pi}\int^{t_0+T/2}_{t_0-T/2} \frac{a^2}{G_4}  \int_{S^2} \int^{\infty}_{-\infty}  \frac{d^3 \rho_{GW}}{df d^2 \Omega} F^P_1(\hat{\Omega}) F_{P2}(\hat{\Omega}) \times \nonumber \\ &\hspace{-150pt} \times  e^{2\pi i f \hat{\Omega}_i (x^i_1 - x^i_2)} f^{-2} df d^2 \Omega dt, \label{SGWB_signal_non_stationary}
\end{align}
the function $G_4$ in the denominator is the generalization of a statistically homogeneous and unpolarized (but possibly anisotropic and non-stationary) SGWB signal in the reduced Horndeski theories. 

The function $G_4$ must vary in the background scale, as discussed near Eq.~(\ref{M_greater_then_R}).
During the period of observation $T \approx 10$ years, these functions can be conceived as constant and so one can simplify
    \begin{align}
      \langle S \rangle = \frac{G_{N}T}{\pi} \frac{1}{G_4(\varphi(t_0))}  \int_{S^2} \int^{\infty}_{-\infty}  \frac{d^3 \rho_{GW}}{df d^2 \Omega} F^P_1(\hat{\Omega}) F_{P2}(\hat{\Omega}) \times \nonumber \\ &\hspace{-140pt} \times  e^{2\pi i f \hat{\Omega}_i (x^i_1 - x^i_2)} f^{-2} df d^2 \Omega, \label{final_product_signal}
\end{align}
where we have set $a(t_0) = 1$. In order to make explicit all factors arising from the Horndeski theory, it is more interesting to express the signal in terms of $N(f, \hat{\Omega})$ using Eq.(\ref{function_N}). We get
\begin{align}
      \langle S \rangle =  \frac{T Q(t_i)}{G_4(\varphi(t_0))}  \int_{S^2} \int^{\infty}_{-\infty}  N(f, \hat{\Omega}) F^P_1(\hat{\Omega}) F_{P2}(\hat{\Omega}) \times \nonumber \\ &\hspace{-140pt} \times  e^{2\pi i f \hat{\Omega}_i (x^i_1 - x^i_2)} f^{2} df d^2 \Omega. \label{signal_with_N}
\end{align}
The function $N$ cannot depend on the modified theory because it is defined by Eq.~(\ref{hom_unp_SGWB}) and we have assumed that the GW Fourier amplitudes $A_{ij}$ are the same as in GR.  Since $\langle S \rangle$ must not depend on the particular instant $t_i$ in which the initial conditions are chosen, we conclude that $Q$ must be independent of $t_i$.   

If we are interested in modified theories of gravity that describe the current accelerated stage of the universe without any change to the GR description of the past, high redshift cosmic epoch, the best way of providing the required initial conditions at $t_i$ is by setting $G_4(\varphi(t_i)) =1$, $a(t_i) = a^{(GR)}(t_i)$ and $h_{ij}(t_i) = h^{(GR)}_{ij}(t_i)$ in the past, so that $Q = 1$. The first of these impositions is an implicit initial condition for $\varphi$ in the high redshift epoch. Once this condition is chosen, $\varphi(t_0)$ will not be arbitrary anymore, but will have a determined value given by the solution of the Friedmann equations in the Horndeski theory, which, assuming the restictions of Eq.(\ref{speed_constraint}), are given by \cite{Kobayashi2011}
\begin{align}
    6H^2G_4 = 16\pi G_N \rho -G_2+ 2X(G_{2,X}-G_{3,\varphi}) \nonumber \\ + 6\dot{\varphi}H(XG_{3,X}- G_{4,\varphi}), \label{Friedmann_I}\\
	2G_4(2\dot{H}+3H^2) = -16\pi G_Np - G_2 + 2X(G_{3,\varphi}-2G_{4,\varphi\varphi}) \nonumber \\ - 4\dot{\varphi}HG_{4,\varphi} +2\ddot{\varphi}(XG_{3,X} - G_{4,\varphi}), \label{Friedmann_II}
\end{align}
where $\rho$ is the matter density and $p$ its pressure.  Thus, it is not trivial that $G_4(\varphi(t_0))$ coincides with its GR value, since $\varphi$ obeys different equations in each theory. We conclude, then, that the ratio $Q(t_i)/G_4(t_0)$ indeed represents a modification of the SGWB signal with respect to GR in this case.

The modification on the SGWB signal can, in principle, be used to probe reduced Horndeski theories by an analogous rationale as used when studying Eq.~(\ref{D_amp}), since in both cases the modification depends on the ratio between $G_4$ evaluated at different events. The difference is that here there is no unique event from which the GW is being emitted. One has to consider the value of $G_4$ today compared with its value at the initial condition hypersurface in the past.

\subsection{The spectral energy density}

When studying the SGWB of astrophysical nature, one usually expresses the spectral energy density appearing in Eq.~(\ref{final_product_signal}) in terms of the luminosity of the galaxies sourcering GWs. Here we briefly argue why the spectral energy density must not change its functional dependence with respect to these luminosities in reduced Horndeski theories.

As a result of purely geometrical arguments, one is able to conclude, for general metric and field equations, that \cite{Cusin2017}:
\begin{align}
	\frac{d^3 \rho_{GW}}{dfd^2\Omega} &=  \int \int \frac{1+z(\vartheta)}{4\pi} \left(\frac{D_A^{(GW)}}{D_L^{(GW)}}\right)^2 \sqrt{p^{\mu}(\vartheta)p_{\mu}(\vartheta)}\times\nonumber \\ &\hspace{30pt} \times n_G(x^{\mu}(\vartheta), \theta_G)\mathcal{L}_G(f_G, \theta_G) d\vartheta d\theta_G, \label{distance_ratio}
\end{align}
where $x^{\mu}(\vartheta)$ is a particular null geodesic arriving at the detector in a given direction and along which gravitons travel, $\theta_G$ is a group of parameters that specify the GW flux emitted by a galaxy, $z$ is the redshift, $n_G$ is the number density of galaxies with parameter $\theta_G$ in a given point of the geodesic, $p^{\mu}$ is the spatial projection of the null geodesic tangent vector, $f_G$ is the frequency of GW as measured near the galaxy and $\mathcal{L}_G$ is the luminosity of a galaxy. Because of what was discussed in Sec.\ref{sec: distances}, the distance-duality relation expressed in Eq.~(\ref{GW_duality}) is still valid in the reduced Horndeski theories and so the ratio of Eq.~(\ref{distance_ratio}) can be replaced by the same explicit function of redshift of the GR case. The quantity $\sqrt{p^{\mu}p_{\mu}}$ can be expressed in terms of redshift as well when one solves the geodesic differential equation in the FLRW metric and uses the Hubble flow observers for the spatial projection. From this, we conclude that the functional dependence of the spectral energy density with respect to luminosity does not change in reduced Horndeski theories. Since, by Eq.~(\ref{function_N}), the spectral GW energy density depends on $Q(t_i)$, we infer such dependency to be present in Eq.~(\ref{distance_ratio}) implicitly, inside the luminosity function.

\section{Conclusion}

The main result of this work is the generalization of the signal of a statistically homogeneous and unpolarized SGWB in a FLRW background metric to the reduced Horndeski theories. It is important to emphasize that no \emph{a priori} relation between the spectral energy density and the SGWB signal was assumed. Such relation was derived from first principles.  {We also took the opportunity to discuss more generically which physical quantities associated to the SGWB differ from their values in GR. Most importantly, we first distinguish between two distances built with GW signals, one defined via its amplitude and another via its flux,  we argue why they must, in general, be different, {we study their relations with the angular diameter distance} and we then clarify their roles in the computation of the energy density of an astrophysical SGWB, that will further impact the measured signal. {Finally, the usual radiation-like dependence with the scale factor is obtained for the energy density of a SGWB, independent of its nature (cosmological or astrophyisical).}}

Eqs.~(\ref{SGWB_signal_non_stationary}), (\ref{final_product_signal}) and (\ref{signal_with_N}) summarize the SGWB signal expressions derived in this work. To simplify the discussion, we neglected the perturbations in the scalar field as well as the longitudinal GW mode. If we want to guarantee that there is a high redshift hypersurface of constant $t$ in which GR is recovered, this must be imposed as an initial condition for the metric and scalar fields. Once such initial conditions are chosen, the scalar field at present time, $\varphi(t_0)$, must, then, be determined using the field Eqs.~(\ref{Friedmann_I}) and (\ref{Friedmann_II}). The additional factor $G_4(\varphi(t_0))$, present in the generalized expressions for the SGWB signal, may be subsequently determined and, thus, does not trivially coincide with its GR value. 

In the context of modified theories of gravity, we showed that there is a conflict between some results found in literature regarding the cosmological distances. If the distance-duality relation between luminosity and diameter distances are valid for both EM and GW signals, as it must be due to photon and graviton number conservation along null geodesics, the luminosity distances cannot coincide with the so called amplitude distance. Once the validity of the distance duality relation for GWs was argued to be true, we were able to conclude that the expression for the spectral energy density of GW per unit solid angle of a SGWB with astrophysical origin has the same functional dependence with the gravitational luminosity of galaxies, Eq.~(\ref{distance_ratio}) together with Eq.(\ref{GW_duality}), as in GR.

Although the EMT of GWs in reduced Horndeski theories, when expressed in terms of the tensorial perturbations of the metric, is generalized to have an additional $G_4$ factor, it was shown that, because there is also a change in the propagation law for the GWs, the SGWB energy density has the same expression in terms of the scale factor as in the GR case, Eq.~(\ref{energy_density_wave_packet}), if we assume that GR is recovered at early times. It is important to notice that, because the Friedmann Eqs.~(\ref{Friedmann_I}) and (\ref{Friedmann_II}) are different from the GR case, the energy density as a function of time can still be numerically distinct from the GR theory, although the functional relation with the scale factor is preserved. 

In this work we assumed that GWs propagate in a FLRW background, and thus the evolution of the background scalar field is dictated by the field equations in cosmological scales, which results in general in a non-constant field. It has been shown, however, that when considering a more general background metric, in the same context of the reduced Horndeski theories, the modifications introduced in the energy momentum tensor or the GW amplitude are also characterized by the factors $G_4(\varphi)$ at specific events \cite{Dalang2020}. In particular, if we were to describe the background considering also the local structures, instead of a purely homogeneous and isotropic cosmology, one would have to take into account screening mechanisms. This raises the issue of whether  such mechanisms play a significant role in the GW signal emitted by both localized sources and the stochastic background. The discussion is essentially about whether the local structures near the emission or the detection events are more important for the signal than the cosmic evolution of the scalar field, in which case one should use the \textit{local} value of $\varphi$.  The value of $G_4(\varphi)$, assuming the local solution for $\varphi$ in the solar system, was already strongly constrained around its GR value by Lunar Laser Ranging (LLR) experiments \cite{Williams2004}. On the other hand it is quite tricky to determine which are the appropriate scales up to which one should detail the background fields, once we are limited by the GW wavelength in one side and by cosmological scales in the other, extreme situations that seem to lead to different results. In view of this discussion, here we simply stress that, a detection of the stochastic background will either probe modified gravity if we work on the cosmological background, or it will constitute a signal insensitive to the theory of gravity (if we assume the LLR bound and that the local gravitational fields are more relevant), giving more model-independent information on the astrophysical side instead.

Possible extensions of our results would be to consider the contribution of the longitudinal mode of GWs to the SGWB signal as well as the role of a scalar wave. If the latter aspect was present, for instance, not only a breathing mode would be expected to appear, but there would be additional contributions to the GW EMT as well as to the propagation Eq.~(\ref{gw_propagation}) (see \cite{Isi2018} for the special case of scalar waves in SGWB for the Brans-Dicke theories). Another aspect worth investigating would be if the SGWB could serve as an alternative probe for constraining the value of the GW propagation speed. This could be investigated by giving up the assumption of Eq.~(\ref{speed_constraint}), which would result in an additional term on Eq.~(\ref{gw_propagation}), changing the amplitude of the wave-packet solution. 

\acknowledgements

We thank Charles Dalang for useful discussions. J. C. L. thanks Brazilian funding agency CAPES for PhD scholarship 88887.492685/2020-00. I S. M. thanks Brazilian funding agency CNPq for PhD scholarship GD 140324/2018-6 and the program CAPES PrInt for scholarship No 88887.569351/2020-0.

\end{document}